\documentclass[conference,10pt]{IEEEtran}
\usepackage{xcolor}
\usepackage{mathtools}
\usepackage{epsfig,makeidx,color,subfigure}
\usepackage{amsmath,amssymb,bbm,enumitem}
\usepackage{cite,graphicx,lipsum}
\usepackage[switch,pagewise]{lineno}
\usepackage{hyperref}
\hypersetup{
        colorlinks = true,
        citecolor=blue,
}
\def\cK{{\cal K}}

\newtheorem{mylemma}{\bf Lemma} 

\def\be{ \begin{equation} }
\def\ee{ \end{equation} }
\def\bea{ \begin{eqnarray} }
\def\eea{ \end{eqnarray} }

\def\b0{{\bf 0}}

\def\cR{{\cal R}}

\def\cS{{\cal S}}
\def\cT{{\cal T}}

\def\sH{{\sf H}}
\def\sI{{\sf I}}

\ifCLASSOPTIONonecolumn
\else

\fi

\usepackage{geometry}
\begin{document}

\IEEEoverridecommandlockouts

\title{Semantic Communication as a Signaling Game with Correlated Knowledge Bases}
\author{
\IEEEauthorblockN{Jinho Choi and Jihong Park\thanks{This research was supported
by the Australian Government through the Australian Research
Council's Discovery Projects funding scheme (DP200100391).}}
\IEEEauthorblockA{School of Information Technology, 
Deakin University\\
Email: \{jinho.choi,\ jihong.park\}@deakin.edu.au}
}

\maketitle
\begin{abstract}
Semantic communication (SC) goes beyond technical communication in which a given sequence of bits or symbols, often referred to as information, is be transmitted reliably over a noisy channel, regardless of its meaning. In SC, conveying the meaning of information becomes important, which requires some sort of agreement between a sender and a receiver through their knowledge bases. In this sense, SC is closely related to a signaling game where  a sender takes an action to send a signal that conveys information to a receiver, while the receiver can interpret the signal and choose a response accordingly. Based on the signaling game, we can build a SC model and characterize the performance in terms of mutual information in this paper. In addition, we show that the conditional mutual information between the instances of the knowledge bases of communicating parties plays a crucial role in improving the performance of SC.
\end{abstract}


\ifCLASSOPTIONonecolumn
\baselineskip 28pt
\fi

\section{Introduction}

Information theory, also known as Shannon’s theory \cite{Shannon}, laid the foundation for modern digital communication technologies and systems such as WiFi, cellular systems, and broadband networks \cite{Proakis08}. According to Shannon's theory, information is characterized as random variables, and efficient compression and transmission schemes through noisy channels can be designed according to their distributions, allowing to analyze their performance limits. While successful, as pointed out by \cite{Weaver53}, Shannon's theory focuses only on the technical problem of accurately transmitting symbols, referred to as \emph{Level A}. This ignores the semantics problem of conveying desired meanings and the effectiveness problem of making the delivered meaning desirable for a given task, namely \emph{Level B} and \emph{Level C}, respectively.


There have been attempts to develop a technical framework reflecting Levels B and C by extending Shannon's theory \cite{Weaver53} \cite{Bao11}. Recently, the desire has been further strengthened by the upcoming sixth-generation (6G) communication systems where semantic communication (SC) is regarded as one of the key enablers \cite{Strinati21}
\cite{Xie21} \cite{Luo22} \cite{CLP_22}. One central issue in developing a framework for SC is how to model the process of mapping between meanings and symbols, i.e., semantic coding. Existing works in the recent literature focus primarily on the algorithmic implementation of semantic coding using a deep neural network \cite{Xie21,dde} that is unfortunately a black-box function without mathematical amenability \cite{achille2019information} \cite{seo2022towards}.

Alternatively, in this paper we aim to develop a theoretic model of semantic coding and thereby propose an analytic SC framework that is compatible with Shannon's theory. To this end, we first focus on the Lewis signaling game \cite{Lewis69} in dynamic Bayesian game theory, where players sequentially make decisions on sending signals in order to convey truthful or meaningful messages \cite{GT_Fudenberg91,Gibbons94,Lewis69}. Achieving the equilibrium of the Lewis signaling game implies the completion of mapping the intended meanings with signals. 

Inspired from this analogy, we formulate the semantic coding through the lens of the Lewis signaling game. Next, we additionally introduce the notion of knowledge bases into the players of the Lewis signaling game. We thereby model how the local knowledge contributes to semantic coding while highlighting the importance of correlated knowledge bases across players. Based on the proposed framework interpreting SC as a Lewis signaling game with correlated knowledge bases, we provide information-theoretic analysis and numerical simulation results, both of which underpin the importance of correlation between knowledge bases. 

Note that a recent study also utilizes a game-theoretic approach to developing an analytic SC model \cite{seo2021semantics} where each player's local knowledge is implicit and assumed to be equivalent to knowing the other players' reactions, as opposed to this work explicitly clarifying the interactions between knowledge and semantic coding. It is also worth noting that recent works \cite{CLP_22, choi2022unified} focus on communicating and synchronizing local knowledge bases, while representing the knowledge using a probabilistic logic language and measuring their amounts of knowledge using semantic entropy \cite{Weaver53,Bao11}. Such knowledge communications can be run on background while communicating signals in the proposed Lewis game-theoretic framework, in order to maintain highly correlated knowledge bases.

\section{System Model}    \label{S:SM}

In this section, we briefly present the Lewis signaling game \cite{Lewis69} and extend it for SC.

\subsection{Lewis Signaling Game} \label{S:LSG}
In the Lewis signaling game \cite{Lewis69}, there are two players, namely the sender and receiver. For convenience, the sender and receiver are called Alice and Bob, respectively.
There are the following three variables:
\begin{itemize}
\item \emph{Types}: $T = t_k$, $k = 1,\ldots, |\cT|$, is a random variable that is observed by Alice.
\item \emph{Signals}: $S = s_m$, $m = 1,\ldots, |\cS|$, is a signal that Alice sends to Bob.
\item \emph{Responses}: $R = r_n$, $n = 1,\ldots, |\cR|$, is a response that Bob chooses.
\end{itemize}
Here, $\cT$, $\cS$, and $\cR$ denote the sets of types, signals, and responses, respectively, having the same cardinaluty $N$, i.e., $N=|\cT|=|\cS|=|\cR|$, which will be relaxed in Sec.~\ref{Sec:LSG_SC}.

\begin{figure}[!t]\vspace{-1em}
\begin{center} 
\subfigure[An optimal equilibrium.]{\label{fig 0 ax}\includegraphics[width=0.22\textwidth]{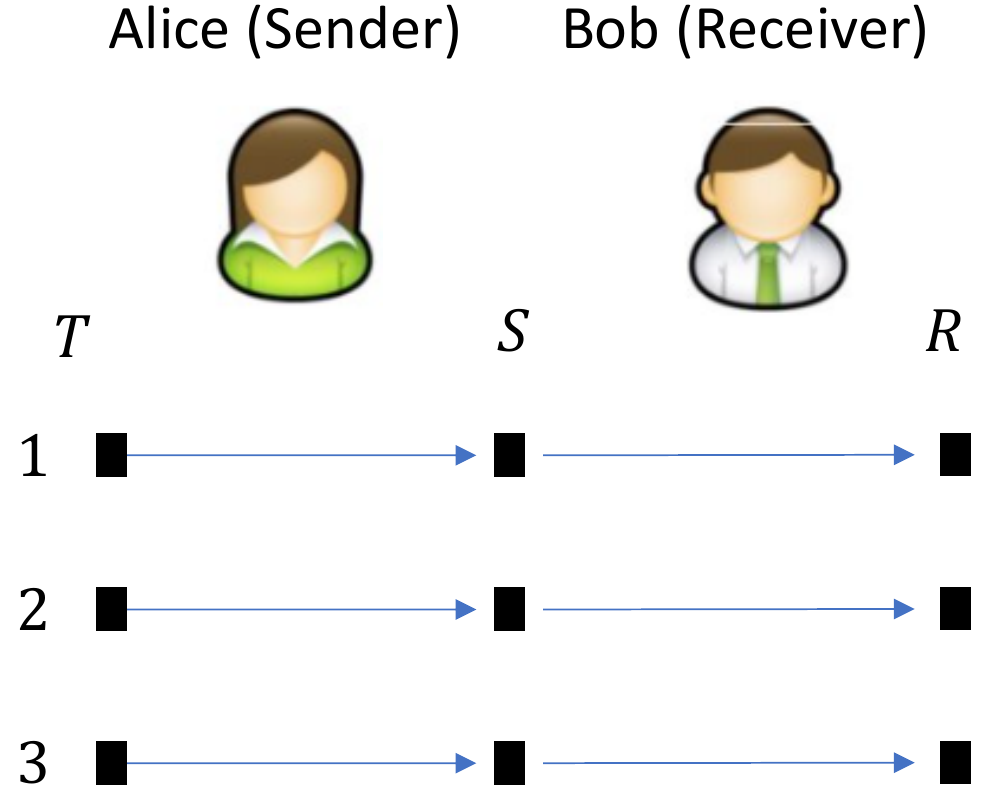}} \hspace{10pt}
\subfigure[A partial pooling equilibrium.]{\label{fig 0 bx}\includegraphics[width=0.22\textwidth]{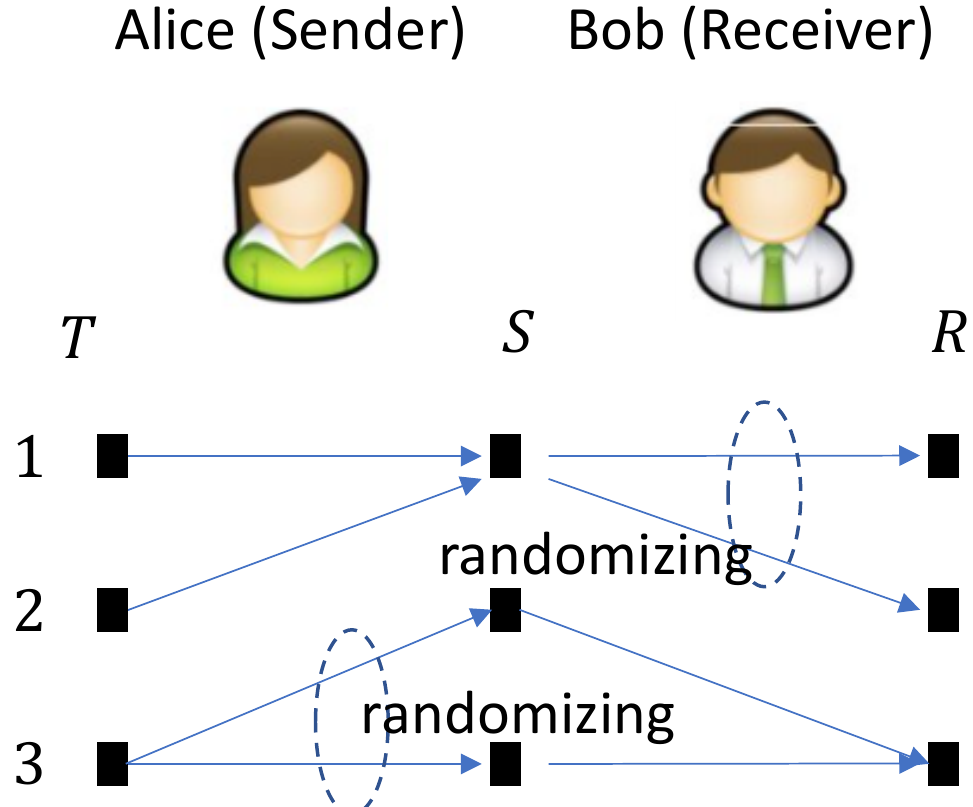}} \vspace{-1em}
\end{center}
\caption{Equilibria of the Lewis signaling game.}
        \label{Fig:eqv}
\end{figure}

In the Lewis signaling game, Alice chooses a signal $S$ to send to Bob, depending on a given type $T$ that is randomly generated from a distribution $\pi$. In this game, Alice moves first, i.e., sending a signal, and then Bob moves next, i.e., receiving the signal and choosing a response, $R$. 
The payoff is given as:
\begin{align} \label{Eq:payoff}
u = \left\{
\begin{array}{ll}
1, & \mbox{if $R = T$;} \cr 
0, & \mbox{otherwise.} \cr 
\end{array} 
\right.
\end{align}
Alice and Bob have mappings. The mapping at Alice is $S = \psi (T)$,
while that at Bob is $R = \phi (S)$. In order to maximize the payoff, Alice and Bob need to choose the mappings such that 
$T = R = \phi (\psi (T))$. There are $N!$ possible sets of mappings to maximize the payoff, including $s_n = \psi(t_n)$ and $r_n = \phi (s_n)$ for $n = 1,\ldots, N$. For example, with $N = 3$, Fig.~\ref{Fig:eqv} (a) visualizes an optimal equilibrium where all the mappings are one-to-one.

There are also undesirable equilibria as shown in  Fig.~\ref{Fig:eqv} (b) where the mappings are not one-to-one. In such a case, the mappings are randomized. For instance, with $N=3$, suppose that each type is generated equally likely, i.e., $\pi (T = t_k) = \frac{1}{3}$ for $k = 1,2,3$. Alice can choose $S = s_1$ if $T = t_1$ or $t_2$, while $S = s_2$ or $s_3$ if $T = t_3$ equally likely, i.e., $\Pr(S = s_2\,|\, T = t_3) = 
\Pr(S= s_3\,|\, T = t_3) = \frac{1}{2}$. Similarly, Bob chooses $R = r_3$ if $S = s_2$ or $s_3$ while $R = r_1$ or $r_2$ if $S = s_1$ equally likely, i.e., $\Pr(R = r_1\,|\, S = s_1) = 
\Pr(R= r_2\,|\, S = s_1) = \frac{1}{2}$. 
In this case, the expected payoff becomes $\frac{2}{3}$. From the given mappings, if Alice or Bob chooses a different mapping, then the expected payoff decreases. In other words, the mappings in Fig.~\ref{Fig:eqv} (b) are in an equilibrium. This undesirable equilibrium associated with randomized mappings are called a partial pooling equilibrium.

Treating the signals as messages (or words) and the types and responses as their intended meanings (or concepts), the Lewis signaling game is akin to the emergence of a language \cite{Catteeuw14} \cite{Gupta21}. From this perspective, the partial pooling equilibrium coincides with the problem of polysemy where a word has multiple meanings. Humans can distinguish the different meanings of a polysemous vocabulary by the aid of their semantic knowledge within a communication context \cite{Grice1975}. Inspired from this, towards modeling SC with machine agents, we extend the original Lewis signaling game by adding the knowledge bases to the agents as we shall discuss in the next section.



\subsection{SC as a Lewis Signaling Game with Knowledge Bases}  \label{Sec:LSG_SC}

In this section we aim to model SC for machine agents by extending the Lewis signaling game described in Sec.~\ref{S:LSG}. To this end, $T \in \cT$ is hereafter referred to as \emph{semantic types} that include semantic information or messages. Alice wishes to deliver $T$ to Bob by sending a signal $S \in \cS$, and Bob chooses its response $R \in \cR$, during which both Alice and Bob utilize their local knowledge bases $\cK_A$ and $\cK_B$, respectively. The objective is to maximize the average of the payoff \eqref{Eq:payoff}, which is now called the \emph{success rate of semantic agreement (SRSA)}. In this SC architecture, we consider the following assumptions \textbf{A1}-\textbf{A3} inspired from human communications, i.e., SC for humans with natural language. 

\begin{itemize} 
\item[{\bf A1})] The number of signals is  smaller than that of semantic types, i.e.,
$|\cS| < |\cT|$.
\end{itemize}

Reducing the number of signals or equivalently maximizing the signaling efficiency is important in coping with the limited number of signaling messages in communication protocols, control signaling, and other promising SC applications \cite{Strinati21,Xie21,CLP_22,Luo22,seo2022towards}. In this respect, it is not preferable to construct signals as many as all possible semantic types. In fact, what humans speak in natural language is much less than what they know and understand. Out of 1.4 million definitions in English, even the Wall Street Journal uses only up to 20k words \cite{paul1992design}. Similarly, emergent machine languages obtained from neural network training often have fewer number of meaningful words as compared with the neural network's maximum expressivity \cite{seo2022towards}. These practical and intuitive motivations rationalize \textbf{A1}.


Under \textbf{A1}, SRSA becomes less than 1. For example, let $|\cS| = 1$ and $|\cT| = N$. Then, Alice only sends $S = s_1$ regardless of types, while Bob can randomly choose one response out of $N$, resulting in $\text{SRSA}=\frac{1}{N}$. This case coincides with the problem of polysemy in natural language where a single word has multiple meanings \cite{Polysemy_Jmemlang}. Humans can overcome this problem by understanding the signals within the communication context based on their local semantic knowledge correlated with each other, e.g., encompassing common general concepts [REF]. Inspired from this, we introduce knowledge bases into Alice and Bob as follows.
\begin{itemize}
\item[{\bf A2})] Alice has her knowledge base $\cK_A$. The instance of her knowledge base at each time, denoted by $K_A\in \cK_A$, affects the generation of semantic type, such that the conditional distribution, $\pi_l (t_k) = \Pr(T= t_k \,|\, K_A = l)$, replaces the distribution of types, $\pi (t_k)$.

\item[{\bf A3})] Bob has his knowledge base $\cK_B$ and its instance is denoted by $K_B \in \cK_B$, which is highly correlated with $K_A$. Bob can choose his response depending on the signal transmitted by Alice as well as $K_B$.
\end{itemize}

In \textbf{A2} and \textbf{A3}, an encyclopedia can be seen as a knowledge base, and its subjects (e.g., technology, agriculture, and so on) are regarded as the instances of the knowledge base. There are multiple items per each subject, which are interpreted as semantic subjects. To illustrate their relationship by an example, suppose that there are two instances of knowledge base, {\tt technology} and {\tt agriculture}. In addition, there are four semantic types, {\tt Apple Computer}, {\tt Raspberry Pi}, {\tt a pack of apples}, and {\tt a pack of raspberries}. For the set of signals,  $\cS = \{\tt apple, \tt raspberry\}$. For signal generation, we have the following mapping:
\begin{align*}
{\tt Apple\ Computer}, {\tt a\ pack\ of\ apples} & \to {\tt apple} \cr
{\tt Raspberry\ Pi}, {\tt a\ pack\ of\ raspberries} & \to {\tt raspberry}.
\end{align*}
At Bob, $S = {\tt apple}$ can be decoded as {\tt Apple Computer} or  {\tt a pack of apples}. To convey the meaning of information more effectively to Bob, Bob needs additional information, e.g., the instance of knowledge at Alice. If Bob knows that the instance of knowledge base used to generate the type is {\tt technology}, he will decode the signal as {\tt Apple Computer}.

Another example is the PatchGame \cite{Gupta21}, a signaling game of referring an intended image while sharing the patch embeddings of the image. Here, the knowledge bases are treated as a set of images, and the image patches are interpreted as the instances of the knowledge base. The image referencing accuracy in the PatchGame increases with the correlation between different patches. Interpreting different patches of an image as the instances $K_A$ and $K_B$, we can expect that SRSA can be improved under correlated $K_A$ and $K_B$, which will be numerically demonstrated in Section~\ref{S:SRSA}.







\begin{figure}[t]
\begin{center}
\includegraphics[width=.35 \textwidth]{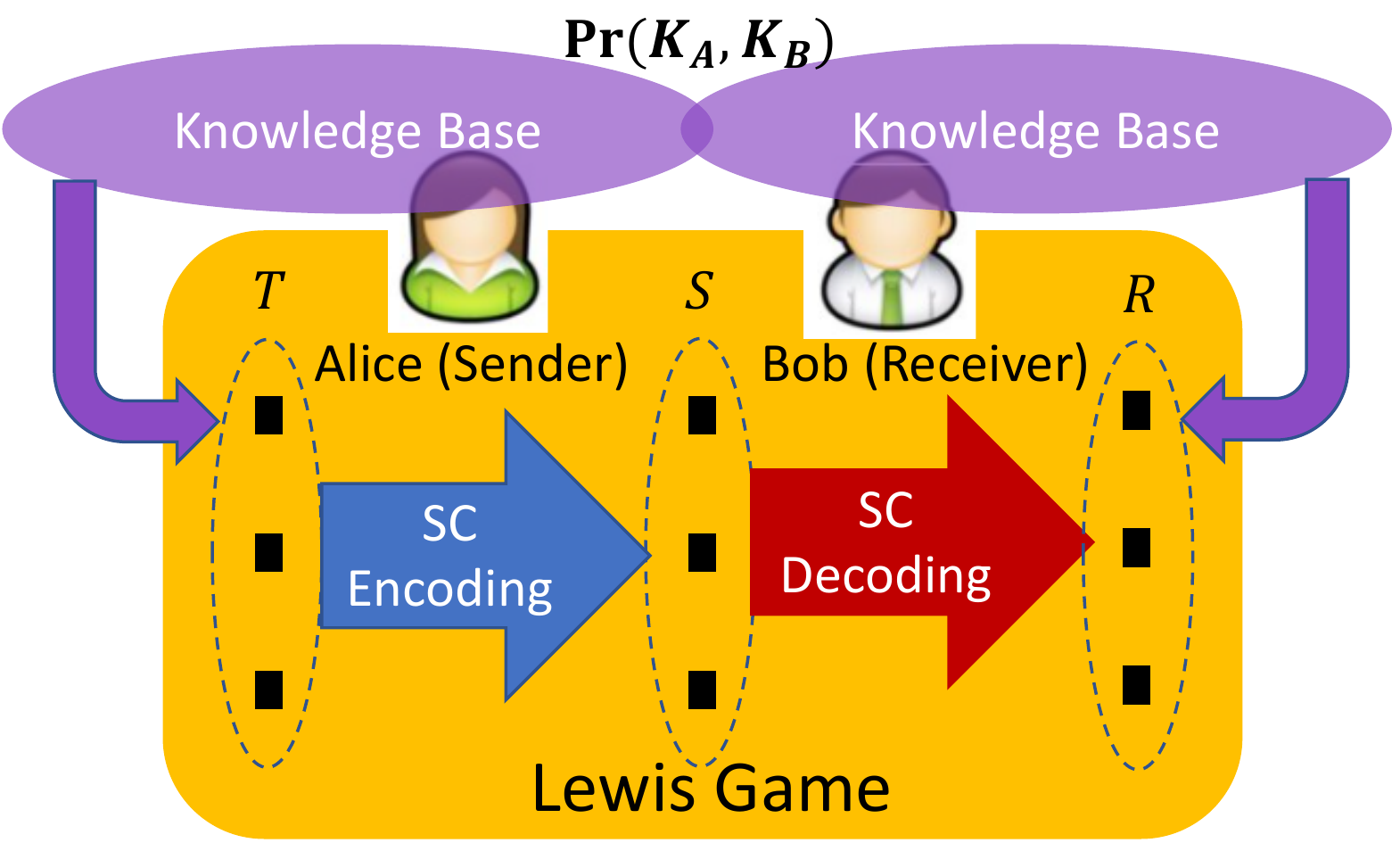} 
\end{center} 
\caption{An SC model based on Lewis signaling game, where $K_A$ and $K_B$ represents the instances of knowledge bases of Alice and Bob, respectively.}
        \label{Fig:a4}
\end{figure}

For mathematical amenability, we hereafter consider that $\cK_A = \cK_B = \cK$ with $|\cK|=L$. We also assume that $|\cR| = |\cT| = N$ and $|\cS| = M \ll N$.
The mapping at Alice for the signal generation is now called SC encoding. The mapping at Bob to decide a response is to be extended,
which is called SC decoding, as follows:
\be 
\phi: \cS \times \cK \to \cR. 
    \label{EQ:SKR} 
\ee 
The resulting SC based on Lewis signaling game is illustrated in Fig.~\ref{Fig:a4}.

\section{Conditions for a High Success Rate of Semantic Agreement} \label{S:SRSA}

For the SC model in Fig.~\ref{Fig:a4}, there are a few key problems:
\begin{enumerate}[label={P}\arabic*)]
    \item What are conditions to achieve a high SRSA?
    \item How to train SC encoding and SC decoding rules for given knowledge bases?
    \item How to build knowledge bases and update for effective SC?
    \item What are fundamental limits of the SC in Fig.~\ref{Fig:a4}?
\end{enumerate}
In this section, we focus on Problems P1 and P4, while P2 and P3 are briefly discussed.

\subsection{Ideal Conditions}

If the instance of Bob's knowledge base,  $K_B$, is independent of that of Alice's, $K_A$, the SC model is reduced to the Lewis signaling game with $|\cS| < |\cT|$. In this case, as discussed earlier, it is difficult to achieve a high SRSA. As a result, for effective SC, it is necessary to impose {\bf A3}, i.e., $K_A$ and $K_B$ are highly correlated. Thus, we first assume that $K_A = K_B$.

Recall that $\psi: \cT \to \cS$ represents the SC encoding at Alice,
while $\phi: \cS \times \cK \to \cR$ in \eqref{EQ:SKR}
the SC decoding at Bob.
\begin{mylemma} \label{L:1}
Under the assumption of $K_A = K_B$, in order to have $R = T$ (for perfect SRSA), with deterministic mappings of $\phi$ and $\psi$, 
a necessary condition is 
\be 
|\cK| \ |\cS| \ (= L M) \ge |\cT| \ (= N).
    \label{EQ:L1}
\ee 
\end{mylemma}
\begin{IEEEproof}
Let $\cT_{(k)}$ denote the subset of the semantic types generated when $K_A = k$, $k = 1, \ldots, |\cK_A|$. For $R = T$ (i.e., Bob is able to choose the semantic type without errors), it is necessary that a pair of the received signal, $S$, and the instance of the knowledge base, $K_B (=K_A)$, should be able to uniquely decide the semantic type, $T$. To this end,
we need
$|\cT_{(k)}| \le |\cS|$, i.e., there should be a sufficient number of signals so that one-to-one mapping from a semantic type 
(within $\cT_{(k)}$) to a signal is possible for a given instance of the knowledge base $K_A = k$. This relation should hold for all $K_A = k$. Thus, since
\begin{align*}
\sum_{k=1}^{|\cK|} |\cT_{(k)}| \ge |\bigcup_k \cT_{(k)}| = |\cT|, 
\end{align*}
we have
$|\cK| \ |\cS| \ge \sum_{k=1}^{|\cK|} |\cT_{(k)}| \ge |\cT|$,
which leads to \eqref{EQ:L1}.
\end{IEEEproof}

As shown in Lemma~\ref{L:1},
provided that \eqref{EQ:L1} holds, 
with deterministic mappings, Bob is able to choose the correct response that corresponds to the semantic type if $K_A = K_B$. This implies that the SRSA becomes 1 if $K_A = K_B$ or 
\be
{\rm SRSA} \ge \Pr(K_A = K_B).
\ee
Thus, for effective SC, it is crucial to have $K_A = K_B$ with a sufficiently high probability. That is, the instances of the knowledge bases of the communicating parties, Alice and Bob, should be highly correlated, while there should be mappings, $\phi$ and $\psi$, 
or SC encoding/decoding rules, that can allow Bob to correctly choose the response under $K_A = K_B$.

\subsection{Information-Theoretic View}

In the context of network information theory \cite{CoverBook}, SC encoding and decoding can be seen as source coding with side information as Bob has his instance of knowledge base, $K_B$, to choose a response. In this subsection, we discuss an information-theoretic view of the SC model in Fig.~\ref{Fig:a4}, which can also be represented by the following Markov chain:
\be
\begin{array}{cccc}
K_A &  \longrightarrow & K_B & \cr 
\downarrow & & \downarrow & \cr 
T & \to S \to & R & (=\hat T)  \cr 
\end{array}
    \label{EQ:Markov}
\ee 
In \eqref{EQ:Markov}, it is clear that Bob choose a response from $S$ as well as side information $K_B$, which can be seen as a noisy version of $K_A$. 

Let $T$ be the source to be transmitted to Bob, while the number of bits that Alice can encode is 
$R_A = \log_2 |\cS|$.
Then, 
Assumption {\bf A1} can be replaced with the following:
\be 
R_A < \sH (T), 
    \label{EQ:RAT}
\ee 
where $\sH(X)$ represents the entropy of random variable $X$.
That is, 
the number of signals is less than the entropy of $T$, which implies that Bob may not be able to correctly choose the response corresponding to Alice's semantic type, $T$, from the received signal, $S$ only. 

Then, at Alice, for SC encoding, it is required that
\be 
R_A \ge \sH(T\,|\, K_A),
    \label{EQ:RATKA}
\ee 
where $\sH(X\,|\, Y)$ denotes the conditional entropy of $X$ for given $Y$. That is, with known $K_A$, the number of signals should be large enough to allow a one-to-one mapping for SC encoding. 
For deterministic mapping in Lemma~\ref{L:1}, \eqref{EQ:RATKA} would be equivalent to 
$|\cS| \ge |\cT_{(k)}|$.
Suppose that each semantic type can be uniquely determined by $K_A$ and $S$, i.e., $T$ and $(K_A, S)$ have a one-to-one mapping, which is referred to as error-free SC encoding at Alice.
In addition, let $R = \hat T = 
\phi (S, K_B)$, where $\hat T$ represents the decoded semantic type at Bob.

\begin{mylemma} \label{L:2}
With error-free SC encoding at Alice,
the mutual information between $T$ and $\hat T$ is given by
\begin{align}
\sI(T; \hat T) = \sH(S) + \sI (K_A; K_B \,|\, S),
    \label{EQ:L2}
\end{align}
where $\sI (X;Y) = \sH(X) - \sH (X\,|\, Y)$ represents the mutual information between $X$ and $Y$, and $\sI (X;Y\,|\, Z)$ denotes the conditional mutual entropy.
\end{mylemma}
\begin{IEEEproof}
Under error-free SC encoding, since
\begin{align*}
f(T \,|\, \hat T) = 
\frac{f (T, \hat T) }{f(\hat T)} = \frac{
f(K_A, K_B, S)}{f(K_B, S)}  = \frac{f(K_A, K_B\,|\, S)}{f(K_B\,|\, S)}, 
\end{align*} 
we have 
\begin{align}  
\sH (T\,|\, \hat T)
& = \sH (K_A, K_B \,|\, S) - \sH (K_B\,|\, S) \cr 
& =\sH (K_A\,|\, K_B, S). 
\end{align} 
Then, it follows
\begin{align}
\sI (T;\hat T) & = \sH(T) - \sH (T\,|\, \hat T) \cr 
& = \sH (K_A, S) - \sH (K_A\,|\, K_B, S) \cr 
& = \sH (S) + \sH (K_A\,|\, S) - \sH (K_A\,|\, K_B, S) \cr 
& = \sH (S) + \sI (K_A; K_B\,|\, S),
\end{align}
which completes the proof.
\end{IEEEproof}

From \eqref{EQ:L2}, we can see that the mutual information between $T$ and $\hat T$ is greater than or equal to the mutual information of $S$, i.e.,
$\sI (T; \hat T) \ge \sH (S)$, since $\sI (K_A; K_B\,|\, S) \ge 0$.
In SC, since the number of bits for signals, $S$, is usually limited, according to \eqref{EQ:L2}, it is important to increase the conditional mutual information between $K_A$ and $K_B$, which depends on the correlation between the two knowledge bases. In other words, with a  limited technical communication bandwidth, the similarity of the two communication parties' knowledge bases, which can be measured by 
the conditional mutual information, $\sI (K_A; K_B\,|\, S)$,
plays a crucial role in SC. As in \cite{CLP_22}, background communication is necessary to synchronize (or correlate) the two knowledge bases as much as possible.

The following result also shows that the SRSA is dependent on the the similarity of the two communication parties' knowledge bases.
\begin{mylemma}
The SRSA is bounded by
\begin{align} 
{\rm SRSA} = \Pr(T= \hat T) 
\le 1 - \frac{\sH (K_A \,|\, K_B, S) - 1}{\log_2 (|\cT|-1)}.
    \label{EQ:L3}
\end{align}
\end{mylemma}
\begin{IEEEproof}
Since \eqref{EQ:L3} can be obtained by applying Fano's inequality, we omit the proof.
\end{IEEEproof}

\section{Simulation Results} 

In this section, we consider SC based on a Lewis signaling game with 
$|\cK| = L = 3$ and $|\cS| = M = 3$. In addition, $|\cT| = N$ is 
$L M = 9$, while $\cT_{(k)} \cap \cT_{(k^\prime)} = \emptyset$,
$k \ne k^\prime$, and $\cup_k \cT_{(k)} = \{1,\ldots, N\}$.
It is also assumed that $\pi_l (t_k) = \frac{1}{M}
= \frac{1}{3}$ for all $l$ and $k$ and $\Pr(K_A = l) = \frac{1}{L}
= \frac{1}{3}$. Thus, each semantic type is chosen equally likely.
For the correlation between $K_A$ and $K_B$, we assume that
\begin{align}
K_B = \left\{ 
\begin{array}{ll}
K_A, & \mbox{with probability $1- \epsilon$} \cr 
U, & \mbox{with probability $\epsilon$,} \cr 
\end{array}
\right. 
\end{align}
where $U \sim {\rm Unif}\{1,L\}$ is an independent random variable. That is, $K_B$ becomes independent of $K_A$ with a probability of $\epsilon$, which is referred to as the error probability of knowledge bases instances.

\begin{figure}[!t]\vspace{-1em}
\begin{center} 
\subfigure[$\epsilon = 0$ ($K_A = K_B$).]{\label{fig 1 ax}\includegraphics[width=0.29\textwidth]{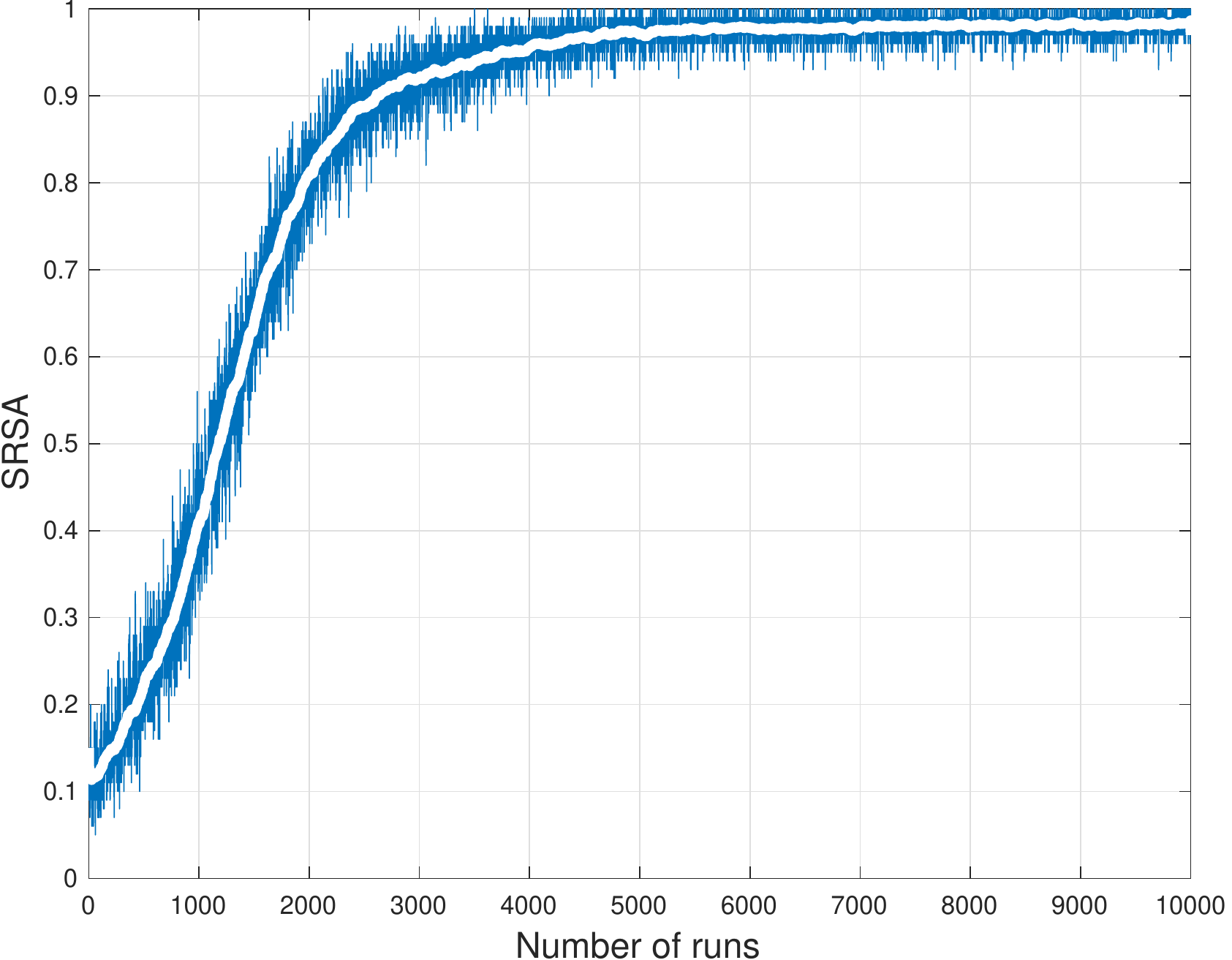}}
\subfigure[$\epsilon = 1$ ($K_A$ and $K_B$ are independent).]{\label{fig 1 bx}\includegraphics[width=0.29\textwidth]{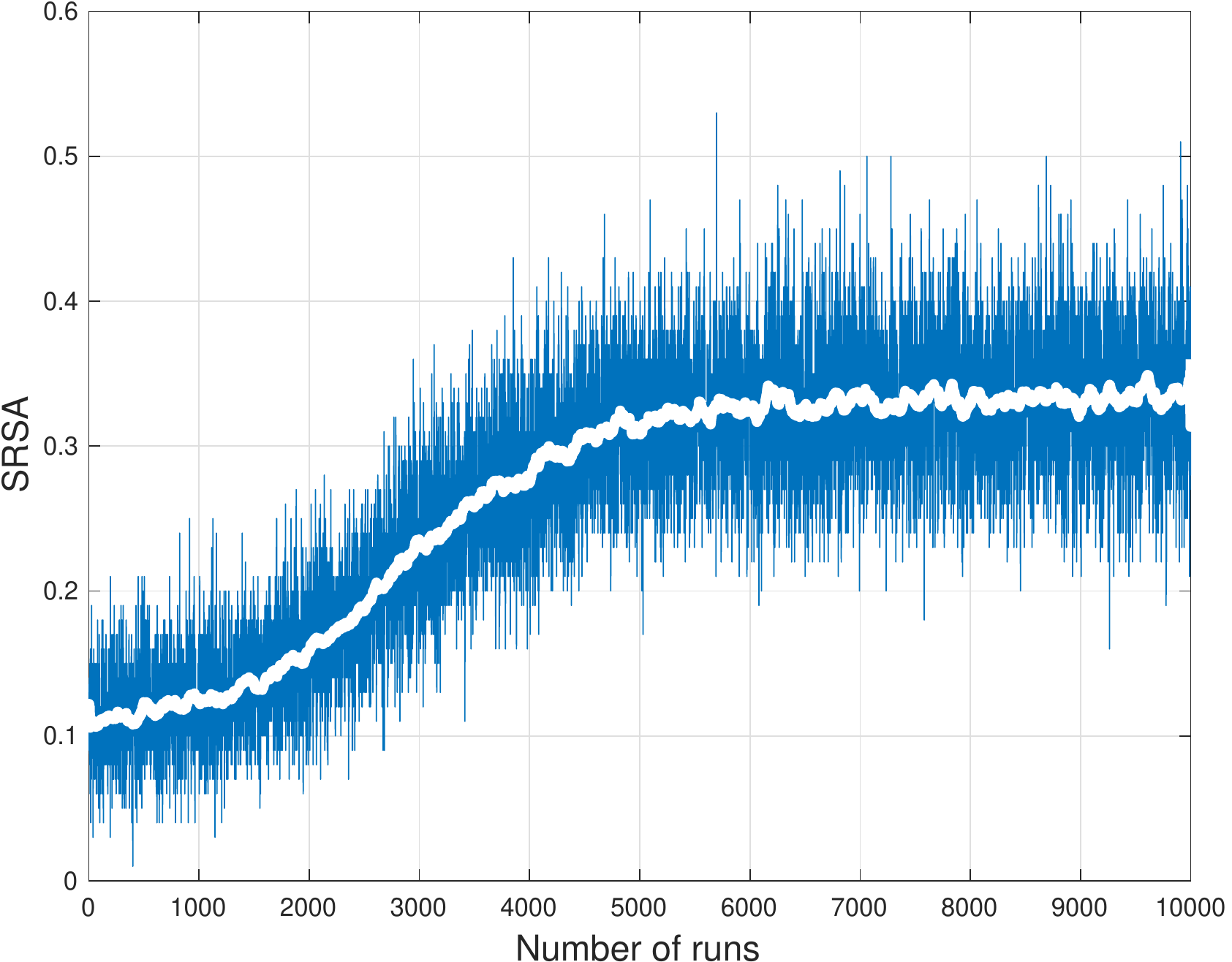}} \vspace{-1em}
\end{center}
\caption{Learning curves for the Q-learning algorithm:
(a) $\epsilon = 0$ ($K_A = K_B$); 
(a) $\epsilon = 1$ ($K_A$ and $K_B$ are independent of each other).}
        \label{Fig:plt_sc}
\end{figure}

For SC encoding and decoding, we use Q-learning \cite{Sutton18}
as in \cite{Catteeuw14} with a learning rate of $0.05$. In Fig.~\ref{Fig:plt_sc}, we show learning curves with $\epsilon = 0$
and $1$. As shown in Fig.~\ref{Fig:plt_sc} (a), with $\epsilon = 0$, the SRSA can approach 1 as Bob's knowledge base instance is the same as Alice's. On the other hand, with $\epsilon = 1$, we can see that the SRSA cannot approach 1 as shown in Fig.~\ref{Fig:plt_sc} (b). Since $K_B$ is independent of $K_A$ and $|\cK| = L = 3$, Bob may have a correct guess of the knowledge base instance with a probability of $\frac{1}{L} = \frac{1}{3}$. Thus, the SRSA can only approach $\frac{1}{3}$.

\begin{figure}[!t]\vspace{-1em}
\begin{center}
\includegraphics[width=.35\textwidth]{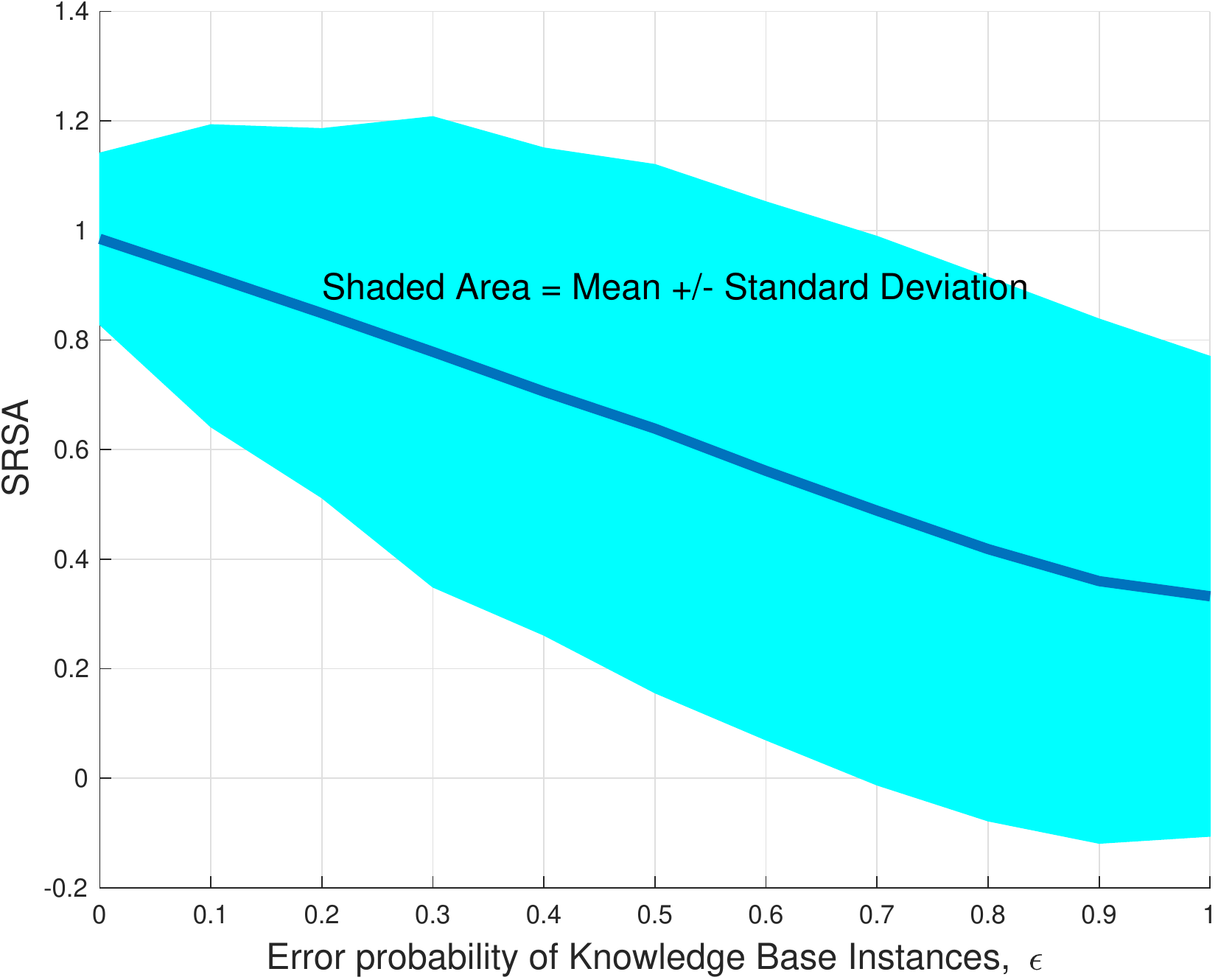} 
\end{center} 
\caption{SRSA curve as a function of $\epsilon$ (note that the shaded area is bounded by the mean $\pm$ standard deviation, while actual values of SRSA cannot be greater than 1).}
        \label{Fig:plt_err}
\end{figure}

The impact of $\epsilon$ on the performance, simulations are carried out with different values of $\epsilon$. As $\epsilon$ increases, the instances of knowledge bases at Alice and Bob disagree with a higher probability. As shown in Fig.~\ref{Fig:plt_err}, for successful SC, it is necessary to ensure that $K_B = K_A$ with a high probability.

\section{Concluding Remarks} 

In this paper, we proposed a SC model based on the Lewis signaling game, where the number of semantic message types is much larger than that of signals. Under this setting, the receiver may not be able to understand the intended message the sender is trying to convey by the signal alone. To address this problem, the knowledge base was used as the source of side information. In particular, it was assumed that the instance of the knowledge base at the sender that affects the current semantic message type is highly correlated with that is available at the receiver, which is to be used to infer the intended message together with the received signal. Based on the proposed SC model, we have been able to determine and characterize the mutual information between the transmitted and received semantic messages at the sender and receiver, respectively. From this result, 
we observed that the conditional mutual information between the instances of the knowledge bases of communicating parties plays a crucial role in 
conveying the meaning of intended messages efficiently
for SC with a limited number of signals (or a limited communication channel capacity). 

In addition to SC, the proposed approach can help characterize the limitation and performance of emergent communication between cooperative neural network or ML agents, which will be considered a topic for further study.

\bibliographystyle{ieeetr}
\bibliography{si}

\end{document}